# Results from the GAMMA experiment on Mt. Aragats

ROMEN MARTIROSOV
*Yerevan Physics Institute, 2 Alikhanyan Br. Str., 0036 Yerevan, Armenia*
ALEKSANDER GARYAKA
*Yerevan Physics Institute, 2 Alikhanyan Br. Str., 0036 Yerevan, Armenia*
SAMVEL TER-ANTONYAN
*Department of Physics, Southern University, Baton Rouge, USA*
ANATOLY ERLYKIN
*P.N.Lebedev Physical Institute of the RAS, Leninsky prospect, 53, Moscow, Russia*
NATALYA NIKOLSKAYA
*P.N.Lebedev Physical Institute of the RAS, Leninsky prospect, 53, Moscow, Russia*
YVES GALLANT
*LPTA, Universit´e Montpellier II, CNRS/IN2P3, Montpellier, France*
LAWRENCE JONES
*Department of Physics, University of Michigan, USA*
JACQUES PROCUREUR
*Centre d'Etudes Nucl´eaires de Bordeaux-Gradignan, Gradignan, France*
HOVHANNES BABAYAN
*State Engineering University of Armenia, 105 Teryan Str., 0105 Yerevan, Armenia*

The present status of the GAMMA facility consisting of an enlarged surface EAS array (116 of 1 $m^2$ scintillation detectors) and underground muon carpet ($150m^2$ detectors) is described. The recent results on mass composition and energy spectrum at the energy region above the knee obtained on the basis of the GAMMA experimental data are presented. It is shown that the power law after the knee is not invariable like -3.1. The slope of the energy spectrum becomes more flat at $E_0$>20 PeV. The strong irregularities of the energy spectrum at about 70-80 PeV are discussed in comparison with other experiments. The bump can be described by a two-component model of primary cosmic ray origin, where additional (pulsar) Fe components are included with a very flat power law energy spectrum.

## 1. INTRODUCTION

The primary goal of many ground-based experiments in cosmic rays is to study the energy spectrum and mass composition of cosmic rays in this energy range. Until recently, the change of the slope of the all-particle energy spectrum from -2.7 below the knee to -3.1 beyond the knee was commonly accepted. Presently there is 30% to 40 % difference in the all-particle energy spectrum obtained from various experiments. A considerably large discrepancy is also observed in experimental data in this energy range on the mass composition of primary cosmic rays. Correct measurement of mass composition can shed light upon the origin of the knee in the energy spectrum. It is necessary to notice that experiments in this energy range are carried out by the ground installations located at various elevations. Therefore, one of the reasons of those discrepancies in experimental data could be due to the large fluctuation of the extensive air showers (EAS) deep in the atmosphere. The other reason might be the difference of the interaction models of the cascade development in the atmosphere.

Special attention should be paid to the energy range 10 - 100 PeV which still lacks in experimental data. At the same time noticeable irregularities of the energy spectrum are observed at these energies. Our group was the first to report this phenomenon in 2002 [Garyaka et al., 2002]. More detailed analysis of this energy range has been presented by us in a more recent publication [Garyaka et al., 2008]. In this work we present experimental results on the energy spectrum and mass composition at $10^7 - 10^8$ GeV derived from the GAMMA experiment on Mt. Aragats in Armenia. The data are improved especially at energies below the knee.

## 2. GAMMA EXPERIMENT

The GAMMA installation was realized in an attempt to study the energy spectrum and mass composition of the primary cosmic radiation in the energy range $10^5 - 10^8$ GeV as well as for investigation of primary very high energy gamma-quanta. The GAMMA (Figure 1) is located on the southern hills of the Mt. Aragats in Armenia with the following geographical coordinates Latitude = 40.47 N, Longitude = 44.18 E and consists of two main parts:

- surface scintillation detectors for registration of the EAS electromagnetic component;
- underground scintillation detectors for registration of the EAS muon component;

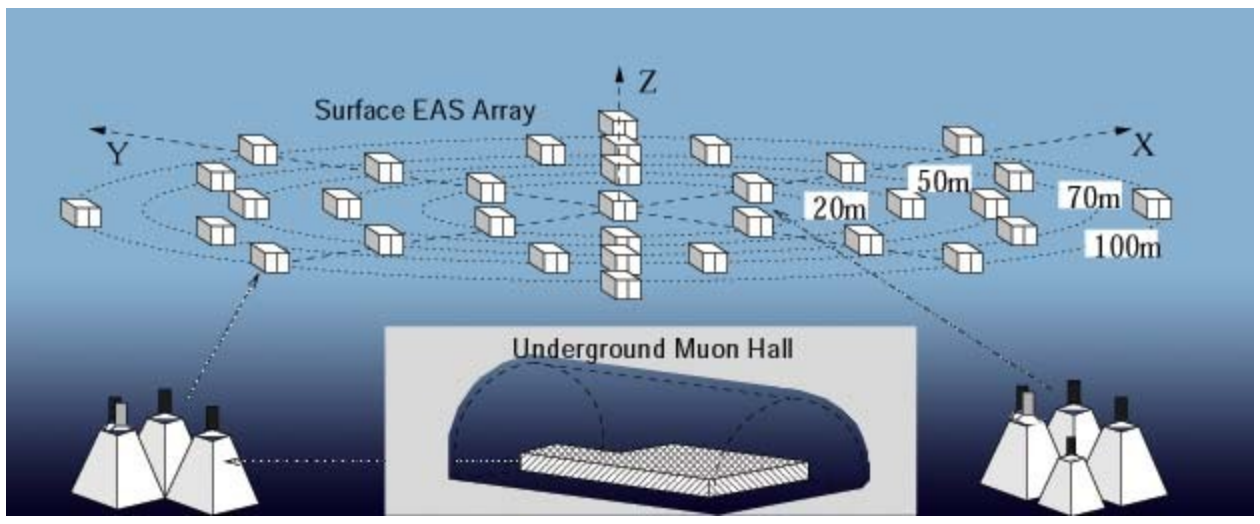

Figure 1. Diagrammatic layout of the GAMMA facility

## 2.1. The surface scintillation array

At present the surface scintillation array consists of 33 groups of three plastic scintillation detectors, arranged in concentric circles with radii of 20, 28, 50, 70 and 100 m. Each detector has an effective area of $1m^2$ and a thickness of $5cm$. The total area of the surface part is about $3\times10^4 m^2$. Each of the central nine stations contains also (4th) small scintillator with dimensions $0.3\times0.3\times0.05m^3$ for high particle density measurements. Recently 8 additional $1m^2$ scintillation detectors were installed at radii 14 and 30 m. It has allowed to reduce the energy threshold up to $10^5$ GeV and to derive more precise data for the primary energy spectrum below the knee.

## 2.2. Underground muon detector

Muon carpet composed of 150 scintillation detectors are compactly arranged in the underground hall under $2.3kg/cm^2$ of concrete and rock. The scintillator dimensions, casings and photomultipliers are the same as in the EAS surface detectors. The arrangement of the muon detectors gives the possibility of determining the muon lateral distribution up to $60m$ at $E_\mu = 5$ GeV.

The EAS angular characteristics (zenith $\theta$ and azimuth $\varphi$ angles) are estimated on the basis of the shower front arrival times measured by the 33 fast-timing surface detectors, applying a maximum likelihood method and the flat-front approach.

# 3. RESULTS

## 3.1. Rigidity-dependent cosmic ray energy spectra [Garyaka et al., 2007]

On the basis of the extensive air shower (EAS) data obtained by the GAMMA experiment, the energy spectra and elemental composition of the primary cosmic rays are derived in the 1-1000 PeV energy range. The reconstruction of the primary energy spectra is carried out using an EAS inverse approach in the framework of the SIBYLL2.1 and QGSJET01 interaction models and the hypothesis of power-law primary energy spectra with rigidity-dependent knees. It is necessary to underline that all the results are derived taking into account the detector response, the reconstruction uncertainties of the EAS parameters, and fluctuations in the EAS development.

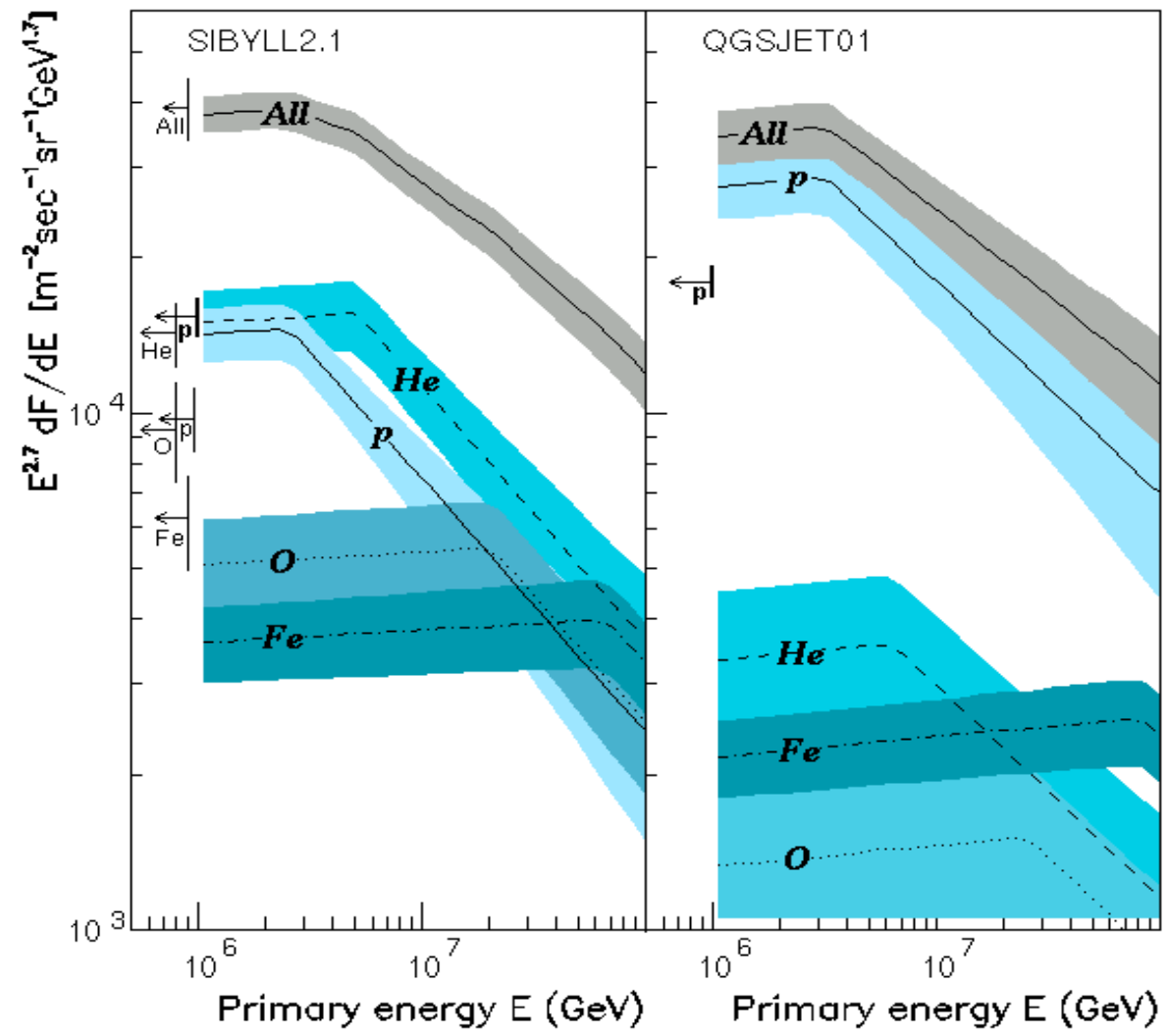

Figure 2. Energy spectra and abundances of the primary nuclei groups (lines and shaded areas) for the SIBYLL (left panel) and QGSJET (right panel) interaction models. All-particle spectra from GAMMA [Ter-Antonyan et al, 2005] and KASCADE [Antony et al., 2005] are shown as symbols. Vertical bars show the extrapolations of balloon and satellite data [Wiebel-Sooth, Biermann&Meyer, 1998].

Energy spectra and abundances of the primary nuclei groups for the SIBYLL and QGSJET interaction models are shown on the Figure 2. As can be seen from this figure, the derived primary energy spectra depend significantly on the interaction model, and slightly on the approach applied to solve the EAS inverse problem. The derived abundances of primary nuclei at an energy $E \approx 1$ PeV in the framework of the SIBYLL model agree (in the range of 1–2 standard errors) with the corresponding extrapolations of the balloon and satellite data [Wiebel-Sooth, Biermann&Meyer, 1998], whereas the results derived with the QGSJET model point toward a dominantly proton primary composition in the 1– 100 PeV energy range.

The corresponding spectral power-law indices are $\gamma_1 = 2.68 \pm 0.02$ and $\gamma_2 = 3.10 - 3.23$ below and above the knee respectively.

## 3.2. All-particle primary energy spectrum in the 3–200 PeV energy range [Garyaka et al., 2008]

Applying a new event-by-event 7 parametric energy evaluation the all-particle energy spectrum in the knee region is obtained on the basis of the data set obtained using the GAMMA EAS array [Garyaka et al., 2007] and a simulated EAS database obtained using the SIBYLL interaction model (Figure 3).

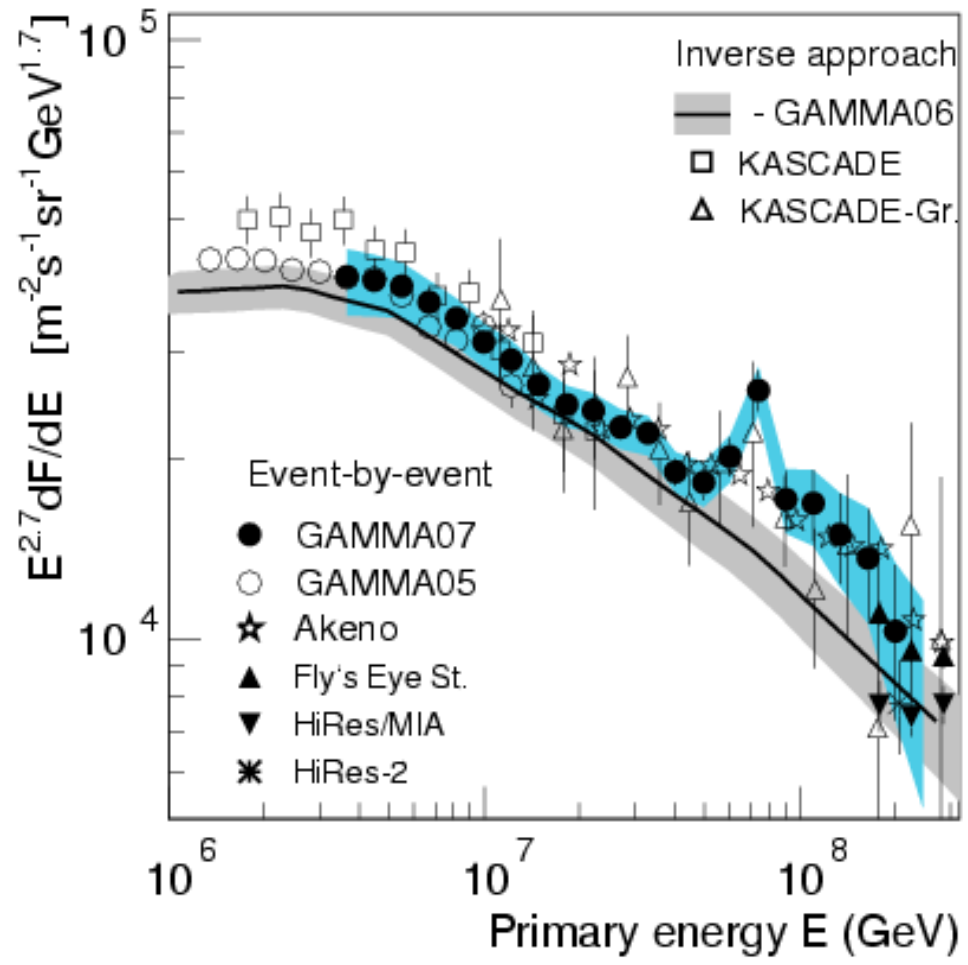

Figure 3. All-particle energy spectrum in comparison with the results of EAS inverse approach [Bruggemann, M. et al, 2006, Garyaka et al., 2007] and our preliminary data [Ter-Antonyan et al, 2005]. The AKENO, Tibet-III, Fly's Eye Stereo, Hires/MIA and Hires-2 data were taken from [Nagano, N. et al., 1984; Amenomori, M. et al., 2008; Bird, D.J. et al., 1995; Abu-Zayyad, T. et al., 2001; Abbassi, R. U. et al., 2002] respectively.

The event-by-event reconstruction of the primary all-particle energy spectrum using the GAMMA facility is mainly based on high correlation of primary energy $E_0$ and shower size ( $N_{ch}$ ). We would like to point out that the power law after the knee is not invariable like -3.1. The slope of energy spectrum becomes more flat at $E_0$>20 PeV. Actually we observed the irregularities in the primary energy spectrum many years ago. This spectrum is shown on Figure 4 [Garyaka, A.P. et al., 2002]. But in that time configuration of surface part of the GAMMA array was more poor and consisted of 75 scintillation detectors located inside of circle with radius 70 m only.

The high accuracy of energy evaluations used in [Garyaka et al., 2008] and small statistical errors point out at the existence of an irregularity ('bump') in the 60–80 PeV primary energy region. According to [Garyaka et al., 2008] the bump can be described by a two-component model of primary cosmic ray origin, where additional (pulsar) Fe components are included with very flat powerlaw energy spectrum $(\gamma_{1p} \approx 1 \pm 0.5)$ before the cut-off energy $E_c, F_e$.

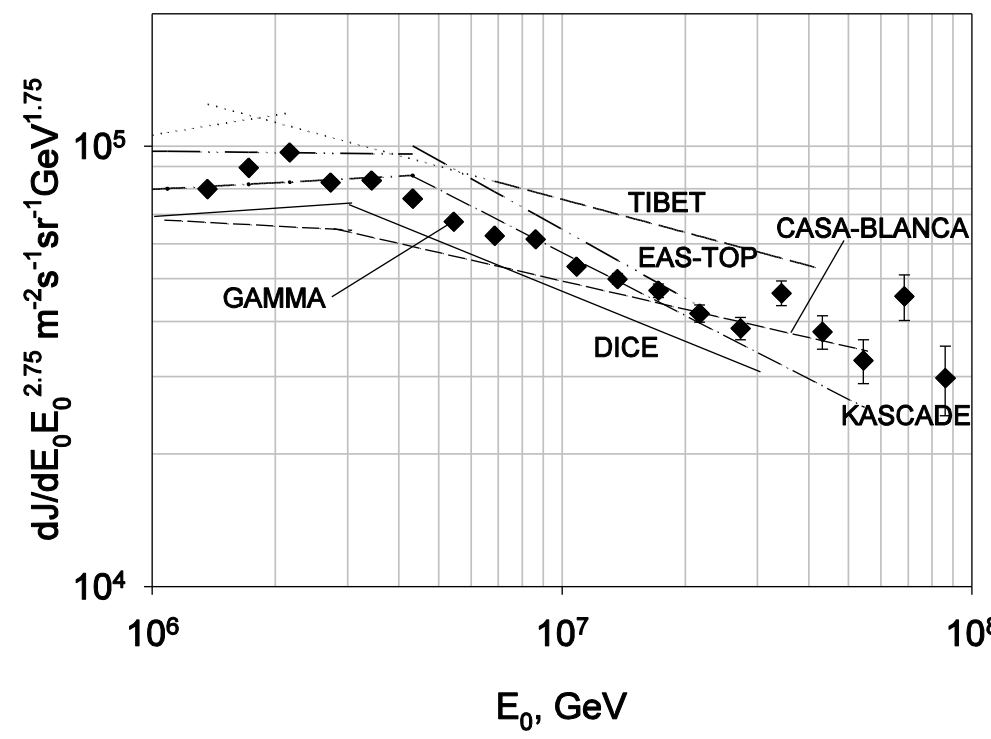

Figure 4 The primary energy spectrum in comparison with results from other experiments. The lines are fitting data of different experiments. The bars are the statistical errors.

Though we cannot reject the stochastic nature of the bump completely, our examination of the systematic uncertainties of the applied method lets us believe that they cannot be responsible for the observed feature.

The indications from other experiments (Figure 5) mentioned in [Erlykin&Wolfendale, 2010] provide the argument for the further study of this interesting energy region.

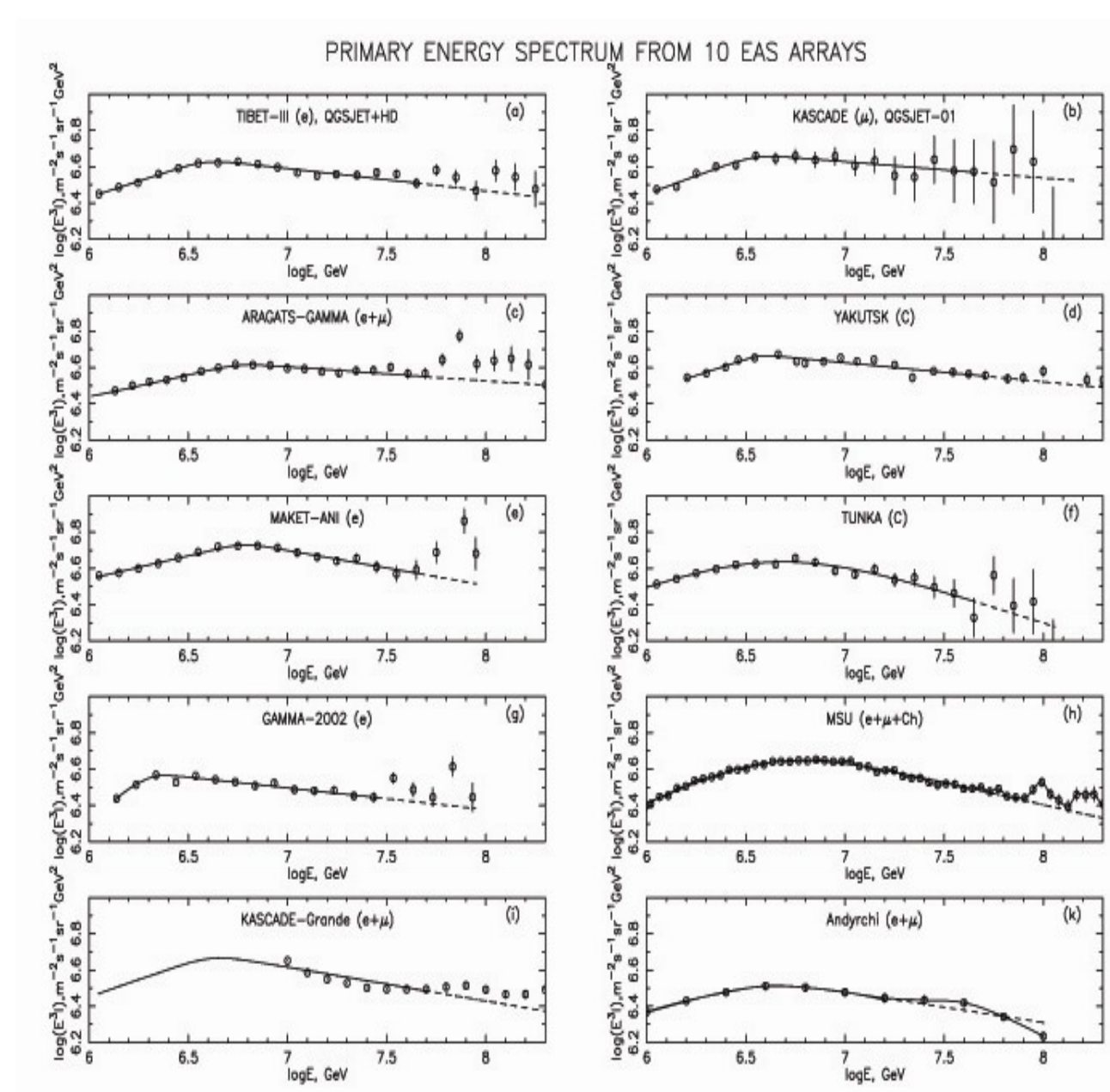

Figure 5. Energy spectra of primary CR, measured by Tibet-III (a), KASCADE (b), GAMMA (c), Yakutsk (d), Maket-Ani (e), Tunka (f), GAMMA-2002 (g), MSU (h), KASCADE-Grande (i) and Andyrchi (k) arrays.